**Electronic, optical, and transport properties of alkali metal oxides (Cs$_2$O): A DFT study**

Anjali Kumari[1], Kamal Kumar[1], Abhishek Kumar Mishra[1]* and Ramesh Sharma[2*]

[1]Department of Physics, Applied Science Cluster, University of Petroleum and Energy Studies, Bidholi via Prem Nagar, Dehradun, Uttarakhand 248007, India

[2]Department of Applied Science, Feroze Gandhi Institute of Engineering and Technology, Raebareli, 229001, Uttar Pradesh, India

## Abstract

The electronic, structural, optical, and thermoelectric properties of the Cs$_2$O cubic structure have been investigated using density functional theory (DFT). The calculations utilize a full relativistic version of the full-potential augmented plane-wave plus local orbitals (FP-APW+lo) method, which is based on density functional theory, employing both the (GGA) and (LDA) approximations. Additionally, we employed the GGA proposed by Trans-Blaha (GGA-mBJ) for band structure computations revealing the indirect band gap nature of Cs$_2$O. The optical properties are also addressed by computing refractive index, extinction coefficient and complex dielectric tensor. The electrical conductivity, Seebeck coefficient, and thermal conductivity exhibit temperature-dependent variations, indicating the formation of a thermoelectric material. Our findings indicate that the compound under investigation is categorized as p-type semiconductor, with the majority charge carriers responsible for conduction being holes rather than electrons.

**\*Corresponding author: mishra_lu@hotmail.com (AKM), sharmadft@gmail.com (RS)**

## 1. Introduction

Alkali metal oxides are crucial for improving the oxidation of various semiconductor surfaces and stimulating catalytic reactions. These oxides are also used for lowering the work function and increasing the electrical current of photocathodes. Because of their high-temperature characteristics, fluorites and antifluorites have gained attention recently. Comparing fluorite ($CaF_2$, $SrF_2$, and $BaF_2$) and antifluorite ($Li_2S$, $Na_2S$, $K_2S$, and $Rb_2S$) materials, abundant research has been done on the former. Under ambient conditions, the alkali metal oxides $K_2O$, $Li_2O$, $Na_2O$, and $Rb_2O$ crystallize in a cubic antifluorite structure (anti-$CaF_2$-type) which is an antimorph of the fluorite structure ($CaF_2$) (#225)[1] These crystals exhibit distinct cation and anion sublattice symmetry. A simple cubic lattice is formed by the anions and an FCC lattice by the cations in the fluorite type-structure. The fluorine ions are located in the calcium tetrahedral structure, whereas the face-centered are occupied by the calcium in the $CaF_2$ structure. Eight fluorine ions envelop each calcium ion, and each fluorine ion is matched with four calcium ions. The anion and cation positions are switched in the antifluorite-type structure, which causes the crystal structure to develop with many huge cavities. The positions of the oxide atoms in the $X_2O$ (X = Li, Na, K, and Rb) compounds are (0; 0; 0) and the X atoms are (0.25; 0.25; 0.25) and (0.75; 0.75; 0.75). Due to metal atoms redistributing on both regular and interstitial sites, these compounds exhibit a wide band gap as well as elevated ionic conductivity. This occurs without causing any significant distortion of the FCC oxide sublattice. This process is known as Frenkel-defect production. Alkali metal oxides are essential for the creation of photocathodes because they stimulate catalytic processes and improve semiconductor surface oxidation [2-3] ]. Therefore, in terms of batteries, solid-state gas detectors or in fuel cells they are promising candidates for these prospective applications.[4], [5]. Hull et al. conducted experiments to test the bulk modulus, lattice parameter and elastic constants of $Li_2O$ at room temperature and up to 1600 K[5] . Alkali metal oxides' electronic structures have been identified by Mikajlo et al. employing both the (LCAO) method and electron momentum spectroscopy measurement[6], [7], [8]. Combining energy loss spectroscopy and photoemission electron- measurements, Liu et al. examined the bulk and surface electronic structure of $Li_2O$[9]. Jupille et al. analysed the oxygen species generated when lithium, potassium, and cesium were present by the use of X-ray and ultraviolet photoelectron spectroscopy studies[10]. Laziki et al. identify the structural phase change of $Li_2O$ from antifluorite to anticotunnite ($PbCl_2$-type structure)[11]. Bidai et al. utilized the ab initio method for determining the elastic characteristics and lattice constants of $Li_2O$, $Rb_2O$ under the effect of temperature and pressure[12]. Cancarevic et al. also used LCAO approach to check the stability of alkali metal oxides under application of pressure [13]. Zhuravlev et al. [14] and Eithiraj and coworkers [1] examined the electronic band structures of the same beneath ambient circumstances utilizing the tight-binding linear muffin-tin orbitals (TB-LMTO) and the self-consistent pseudopotential method (PP), respectively. Utilizing the molecular dynamics (MD) simulation method, the superionic behavior of $Li_2O$ was studied. The elastic constants and the lattice constant under ambient pressure were determined from the simulation[1], [15]. Wilson et al. estimated the structural and elastic properties of lithium oxides[16]. The asphyrical ion model and the LDA-pseudopotential plane wave approach were employed by the several other authors. Utilizing both the LCAO and plane wave (PW) approaches, Islam et al. investigated the structural, electrical, and defect characteristics of lithium oxide [17]. Mauchamp and associates employed the full potential linearized augmented plane wave (FP-LAPW) technique to mimic the near-edge structure electron energy-loss at the lithium K edge in lithium oxide[18]. Understanding the optical and electrical characteristics of materials under strain offers valuable perspectives on how well they will function in real-world scenarios. The valence bandwidth and lattice parameters of lithium, sodium and potassium oxides are predicted using the linear combination of atomic orbitals by Hartree-Fock (LCAO-HF) method and DFT research[15] and the computed valence bandwidth was then compared to the findings of electron momentum spectroscopy. Alkali metal oxides and sulphides optical band gap was covered by Zhuravlev et al. [14] . For $Li_2O$ and $Na_2O$, the Wannier function-based HF technique has been published; this investigation highlights the significance of correlation effects. Only valence bandwidth investigations have been conducted for $Li_2O$, $Na_2O$, and $K_2O$, and no electronic band structure calculation has been reported for $Rb_2O$ so far. Hence, it is appropriate to calculate the properties of $Cs_2O$ compounds in order to gain insights on theoretical works on this fascinating class of materials and to provide reference data for the experimentalist. In this study we investigated the structural, electronic, optical, and thermoelectric properties of $Cs_2O$. To the best of our knowledge, no previous research has been done on this compound.

## 2. Computational Methodology

The $Cs_2O$ material was simulated by utilizing the Full Potential Linearized Augmented Plane Wave (FP-LAPW) method, implemented through the WIEN2k package[19]. To achieve more accurate results for band gaps and lattice constants, first, we used the generalized gradient approximation (GGA), specifically the Perdew-Burke-Ernzerhof (PBE) functional, for calculations of the exchange-correlation energy[20], [21]. This approach captured the overall energetics effectively. However, to achieve high accuracy for the bandgap prediction, we then employed the Tran-Blaha modified Becke-Johnson (TB-mBJ) exchange-correlation functional. This more sophisticated approach comes at a higher computational cost, but it provides the necessary level of detail for accurate bandgap calculations. Thermoelectric properties were evaluated using the BoltzTrap code[22]. The cut-off parameters were set to $RMT \times Kmax = 8$, $Gmax = 14$, and $Length = 10$, with an energy threshold of 9.0 eV between core and valence states. Self-consistent calculation convergence was determined when the system stabilized within $10^{-5}$ eV for total energy and force. To integrate over the Brillouin zone, we used the Monkhorst-Pack method with 18 special k-points within the irreducible Brillouin zone[23], [24].

## 3. RESULTS AND DISCUSSION

### 3.1 Structural properties

In the $Cs_2O$ compounds, the oxide atoms are located at (0; 0; 0) and the Cs atoms are located at (0.25; 0.25; 0.25) and (0.75; 0.75; 0.75) positions. When the total energies versus lattice parameters are mapped to the Murnaghan equation of state [39], the equilibrium lattice parameter (a0), bulk modulus B0, and its pressure derivative B are derived.

$$E(V) = E_0 + \frac{9V_oB_o}{16}\left[\left\{\left(\frac{V_o}{V}\right)^{2/3} - 1\right\}^2 B' + \left\{\left(\frac{V_o}{V}\right)^{2/3} - 1\right\}^2 \left\{6 - 4\left(\frac{V_o}{V}\right)^{2/3}\right\}\right] \quad (1)$$

Our study explores the structural properties of $Cs_2O$ at zero pressure and temperature using Density Functional Theory (DFT) calculations. Table 1 summarizes the calculated lattice constant, bulk modulus, and its pressure derivative. The results show good agreement with previous theoretical investigations by other researchers [13], [25].When compared to experimental data, a well-known trend emerges with the choice of functional. The Generalized Gradient Approximation (GGA) slightly overestimates the lattice constant of $Cs_2O$, while the Local Density Approximation (LDA) underestimates it. This aligns with the general behavior of these functionals. The calculated bulk modulus of $Cs_2O$ suggests that $Cs_2O$ can deform more readily under pressure. To ensure the accuracy of our findings, we performed convergence tests using various muffin-tin radii and k-point sets. This meticulous approach guarantees the reliability of the calculated structural properties. It's important to note that this study goes beyond structural properties. We have also calculated other physical characteristics of $Cs_2O$, including electronic, optical, and thermoelectric properties. These additional calculations provide a more comprehensive understanding of the material's behavior.

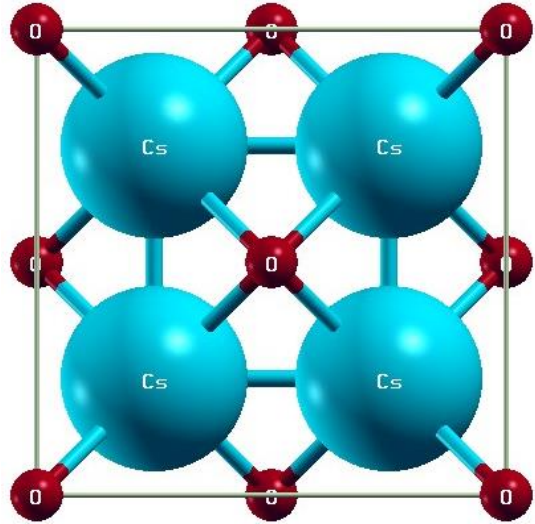

Fig. 1 : Conventional unit cell of the cubic $Cs_2O$

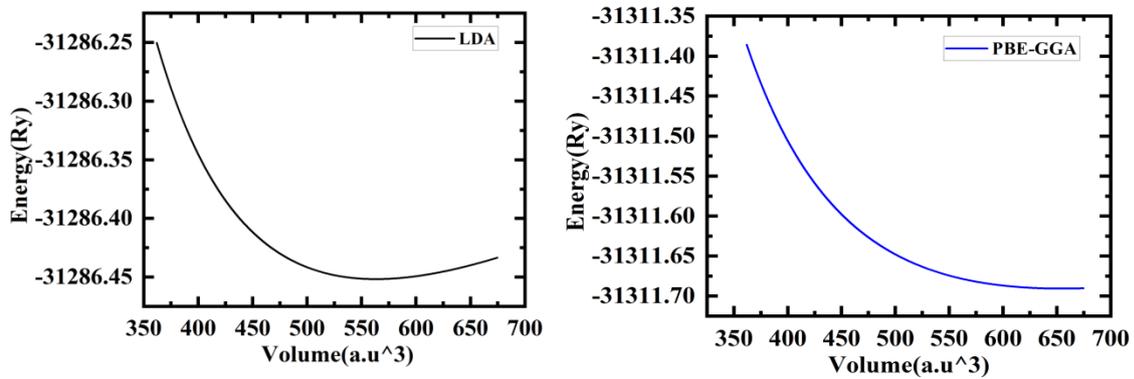

Fig 2 : Energy versus volume of $Cs_2O$

To determine the structural stability of the $Cs_2O$ cohesive energy ($E_{Coh}$) was calculated. The equation below demonstrates the cohesive energy per formula unit cell.

$$E_{Cohessive\ energy} = xE_{Cs}^{iso} + yE_{O}^{iso} - E_{tot}^{Cs_2O} \qquad (2)$$

where $E_{tot}^{Cs_2O}$ being the total energy of $Cs_2O$ in the ground state and $E_{Cs}^{iso}$ and $E_{O}^{iso}$ are the individual atoms energies i.e O and Cs of the compound where x, y, and z are concentrations of the elements. The elements in Cs₂O alloys have the large values of cohesive energy which is positive (3.45 eV/f.u.c) as shown in Table 2. The decrease in $E_{Coh}$ due to the increase in the atomic radius of the Cs₂O materials can be seen in Table 2. An additional to confirm the stability of the Cs₂O is to compute the formation ($E_{for}$) energy from Equation (3) as follows:

$$E_{Formation\ energy} = E_{tot}^{Cs_2O} - [xE_{bulk}^{Cs} + yE_{bulk}^{O}] \tag{3}$$

where $E_{tot}^{Rb_3OX}$ is the total formation energy and the rest of the quantities, $E_{bulk}^{Cs}$ and $E_{bulk}^{O}$ are the energies of individual atoms. The formation energy of the Cs₂O materials is also shown in Table 1. A negative formation energy of a compound i.e., -1.01 is considered as thermodynamically stable. From table 1 it can be understand that the Cs₂O has a negative formation energy which shows that it is thermodynamically stable. We found a satisfactory match between our projected formation energies for the inverse Cs₂O alloys and the alkali oxide compounds found in the materials project data[26].

## 3.2 Electronic properties

The self-consistent full relativistic band structures of Cs₂O in the cubic phase at equilibrium volume were obtained using LDA, PBE-GGA, and mBJ-GGA schemes. The LDA and PBE-GGA generally shows underestimation of energy gaps because their simple forms lack the flexibility needed to accurately reproduce both the exchange-correlation energy and the charge derivative. To address this shortcoming, Trans-Blaha developed a new GGA functional designed to improve the exchange-correlation potential, although it provides less accurate exchange energy. This new approach, called mBJ-GGA, has been shown to yield better results for quantities dependent on the energy eigenvalues, including band gaps.

The computed band structures using GGA, LDA, and mBJ-GGA for Cs₂O were almost similar, except for the band gap values, which were higher with mBJ-GGA. Fig 3. Shows the bandgap plots for the compound. The valence band maximum (VBM) is located at the X point, and the conduction band minimum (CBM) is at the Γ point, indicating an indirect band gap for Cs₂O.

The calculated band gaps using GGA, LDA and mBJ approximations are presented in Table 2 which clearly shows that the mBJ- calculated band gaps significantly improve over the LDA and PBE-GGA values.

Additionally, to understand the electronic band structure, the total and partial densities of states (DOS) were calculated, as displayed in Fig. 4 (a, b). Our results are consistent with those obtained by Eithiraj et al. Fig. 3 clearly shows that $Cs_2O$ is an indirect band gap (X-Γ) material. Additionally, the TDOS and PDOS calculations reveal the participation of various states in forming energy bands, with the TDOS graph indicating that the energy states in the valence band range from −15 eV to the Fermi level. The conduction band edge is located at 5 eV for $Cs_2O$. A detailed examination of the PDOS graph shows that the formation of the valence and conduction bands is significantly influenced by the 'p' states of O and the 'p' and 'd' states of Cs.

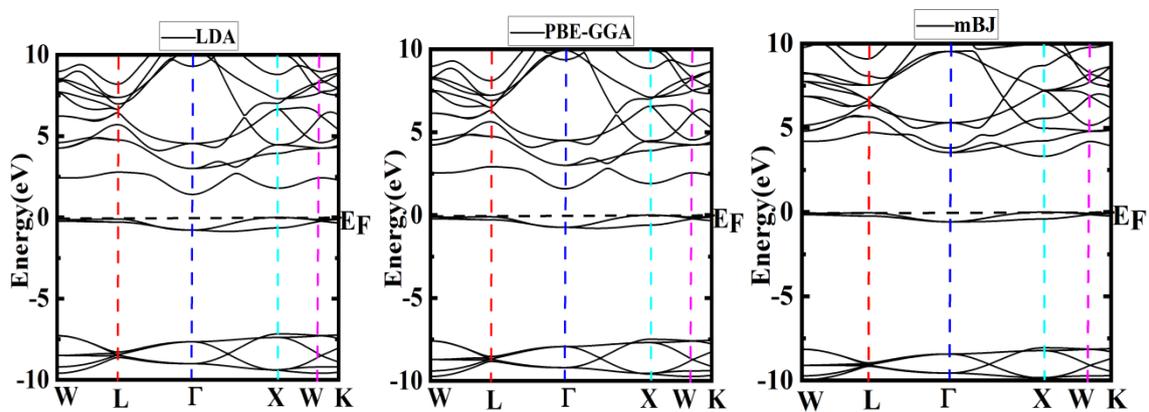

Fig.3 Band structure at equilibrium lattice constant for $Cs_2O$

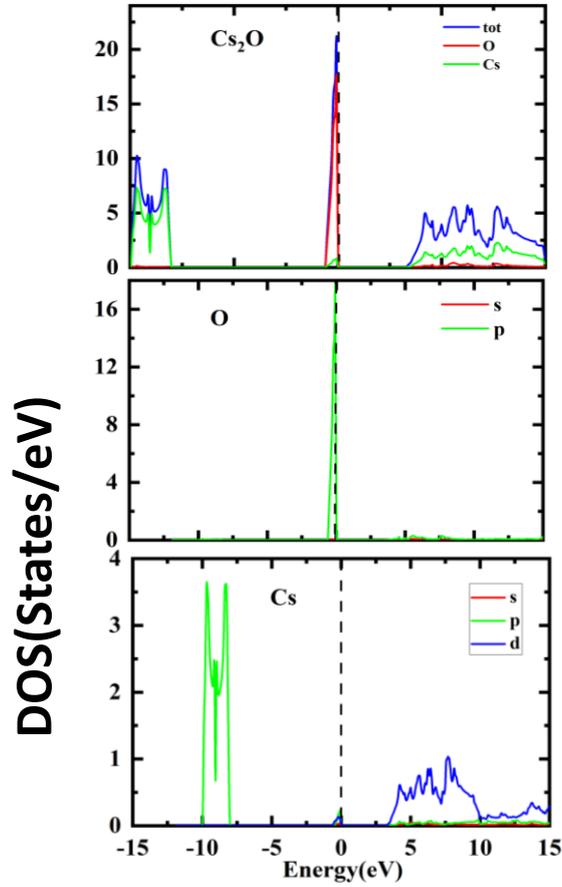

Fig.4 Total and Partial density of states (pDOS) at equilibrium lattice constant for Cs$_2$O

**Charge density**

In order to understand the nature of chemical bonding and atomic interactions between the atoms in the Cs$_2$O, we have calculated the charge density plots presented in Fig. 5. The vertical color bar on the left side indicates the total electron density, ranging from 0 (bottom) to 1 (top), representing the degree of electron localization. Red areas denote regions with highly localized electrons, while blue areas indicate zero localization. The charge density around the nuclei of the ions in Cs2O displays a nearly spherical distribution, similar to that of noble gas atoms. The larger size of Cs atoms compared to O atoms is evident. Additionally, the charge transfer from the electropositive Cs atoms to the nearby oxygen and halogen atoms, due to their higher electronegativity, is observable. The distorted outer shell of Cs ions indicates charge transfer from Cs to O, as oxygen has the highest electronegativity (3.5). The spherical charge density distribution around the softer oxygen and halogen elements is more pronounced than around the harder alkali metal Cs atoms. Therefore, our DFT investigation suggests that inverse Cs2O materials are ionic, characterized by electron transfer from the electropositive Cs to the electronegative oxygen, a hallmark of ionic materials. The charge density plots of Cs2O in the (110) plane, shown in Fig. 3, confirm the material's dominant ionic character.

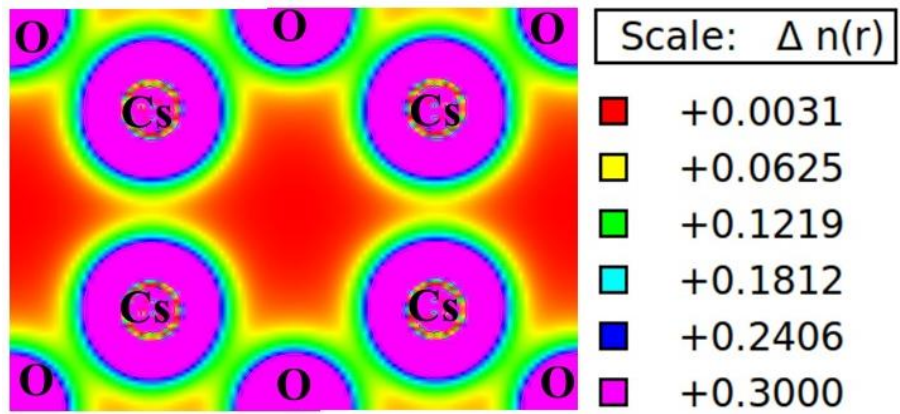

Fig.5 The charge density plots for Cs$_2$O

### 3.3. Optical properties

To fully define the linear optical characteristics, we only calculated single dielectric tensor component because the alkali metal oxides have cubic symmetry. Here, the frequency-dependent dielectric function is represented by ε(ω). The Kramers–Kroning relation determines the real part of the dielectric constant [27], [28].

$$\varepsilon_1(\omega) = 1 + \frac{2}{\pi} P \int_0^\infty \frac{\omega' \varepsilon_2(\omega')}{\omega'^2 - \omega^2} d\omega' \qquad (4)$$

P is illustrating the principal integral

$$\varepsilon_2(\omega) = \frac{e^2 \hbar}{\pi m^2} \sum_{v,c} \int_{BZ}^\infty |M_{CV}(k)|^2 \, \delta[\omega_{cv}(k) - \omega] d^3 k \qquad (5)$$

We can calculate the n (ω), R (ω), and, a (ω) by interpreting of $\varepsilon_1(\omega)$ and $\varepsilon_2(\omega)$. We can explore these relationships by employing the equations provided.

$$R(\omega) = \frac{[n-1]^2 + k^2}{[n+1]^2 + k^2} \tag{6}$$

$$n(\omega) = \sqrt{\frac{\{\varepsilon_1^{\ 2}(\omega) + \varepsilon_2^{\ 2}(\omega) + \varepsilon_1(\omega)\}}{2}} \tag{7}$$

In Fig. 6 (a), both the real and imaginary components of the dielectric function are depicted. Given the cubic structure of the compound, a single component is adequate to fully describe its optical properties. In the lower energy range, the material displays transparency, as evidenced by the imaginary part of the dielectric function. An absorption edge is observed at 3.0 eV, indicating the first electron transition from the (VBM) to (CBM). The primary spectral feature is centered around 7.0 eV, accompanied by additional peaks corresponding to electron transitions from lower to higher energy bands.

The Kramers-Kronig relations is used in determining the real part, $\varepsilon_1(\omega)$, of the frequency-dependent dielectric function from the imaginary part, where P denotes the principal value of the integral. And understanding both the parts enables us to calculate crucial optical functions such as the refractive index, among others.

$$\varepsilon_1(\omega) = 1 + \frac{2}{\pi} P \int_0^\infty \frac{\omega' \varepsilon_2(\omega')}{\omega'^2 - \omega^2} d\omega' \tag{8}$$

$$\varepsilon_2(\omega) = \frac{e^2 \hbar}{\pi m^2} \Sigma_{v,c} \int_{BZ}^\infty |M_{CV}(k)|^2\ \delta[\omega_{cv}(k) - \omega] d^3k \tag{9}$$

$$R(\omega) = \frac{[n-1]^2 + k^2}{[n+1]^2 + k^2} \tag{10}$$

$$n(\omega) = \sqrt{\frac{\{\varepsilon_1^{\ 2}(\omega) + \varepsilon_2^{\ 2}(\omega) + \varepsilon_1(\omega)\}}{2}} \tag{11}$$

The real part of the dielectric, depicted by the blue line in Fig. 6 (a), exhibits a maximum peak at 3.0 eV, gradually decreasing with increasing photon energy and reaching a minimum value at 6.9 eV.

Fig. 6 (d) illustrates the absorption coefficient α(ω) plot concerning photon energy (eV). This coefficient is utilized to gauge the attenuation of photon energy within a material medium. As evident from both Fig. 6(a) and Fig. 6(d), the absorption coefficient and the imaginary part of the dielectric constant follow similar trends, as they both describe light absorption. With increasing photon energy, α(ω) rises to a maximum value of 80 x 10$^4$/cm at 6.5 eV. This sharp increase commensurates to the limit of incident photons reaching the absorption edge. Subsequently there's an increase in photon energy with a decrease in absorption coefficient, indicating the semiconducting nature of the computed material.

Moreover, through light reflection, the materials' surface morphology under analysis can be examined, denoted as R(ω), as shown in Fig. 6 (e). R(ω) values calculated at zero photon energy were approximately 0.05 (refer to Table 2), negligible in impact. As photon energy increases, R(ω) rises to a maximum value of 0.25 at 6.5 eV. Such low R(ω) values have minimal effect on the performance of optical devices.

Another critical parameter to consider is the energy loss function, denoted as L(ω), encompassing intra-band, inter-band and plasmonic interactions. Additionally, optical loss serves as a metric for energy dissipation through scattering and dispersion as illustrated in Fig. 6(f).

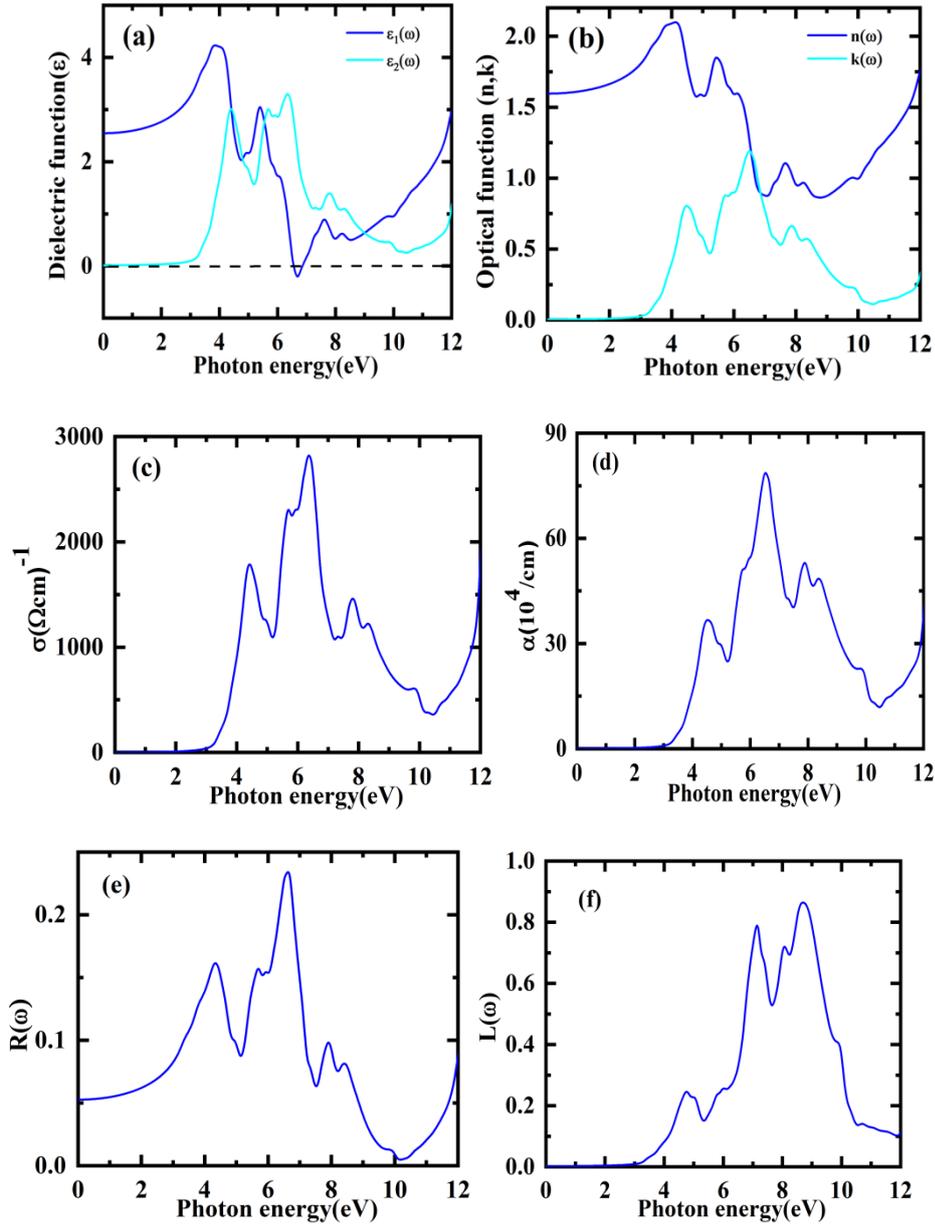

Fig. 6: Calculated optical properties plot for Real and Imaginary parts

## 3.4 Thermoelectric properties

Scientists are becoming more and more interested in thermoelectric (TE) generators because of their exceptional efficiency in transforming heat energy into electrical energy. TE materials are used in computer cooling systems, TE refrigeration, and small detector components, among other uses at different scales. When comparing various temperature levels to characteristics such as power factor, Seebeck coefficient, thermal and electrical conductivity, they are especially useful. These metrics are important markers that offer important information about the thermoelectric responses of materials at different temperatures. To evaluate a material's behavior in high-temperature settings, one must have a thorough understanding of its thermoelectric properties. Semiconductors with narrow bandgap values are recognized for their effectiveness as thermoelectric materials. By analyzing band structures, electronic transport coefficients are evaluated using a semi-classical generalized BoltzTrap program and a rigid band approximation. In this study, thermoelectric properties are computed over a temperature range of 0–1200 K.

The temperature-dependent electrical conductivity and Seebeck coefficient can be expressed as[29], [30]:

$$\sigma_{\alpha\beta}(\alpha,\mu) = \frac{1}{\Omega}\int \sigma_{\alpha\beta}(\varepsilon)\left[-\frac{\partial f_0(T,\varepsilon,\mu)}{\partial \varepsilon}\right]d\varepsilon \qquad (12)$$

Where $\sigma_{\alpha\beta}(\varepsilon)$, f, $\Omega$ and $\mu$ are transport energy distribution tensor, Fermi-distribution function, unit cell volume, and chemical potential, respectively. The variables α and β represents the tensor indices of transport distribution. Electrical conductivity measures free charge carrier flows.

**Seebeck Coefficient**

The temperature dependence plot of calculated Seebeck coefficient of $Cs_2O$ has been presented in Fig. 7 (a). From the plots S values of $Cs_2O$ remain positive throughout the specified temperature range which suggests that p-type charger carrier are dominant here. The value of S at 100K is 280($\mu V/K$). As temperature increases the value of S decreases.

**Electrical Conductivity**

Fig. 7(b) shows how electrical conductivity over the constant relaxation time (σ/τ) varies with temperature in the range of 100–1200 K. Since the relaxation time (τ) is constant and independent of temperature, the (σ/τ) ratio increases with rising temperature, peaking at a value of 2.5x1020 (Ω m s) $^{-1}$, indicating semiconducting behaviour.

**Thermal Conductivity**

Another crucial parameter for evaluating the thermoelectric performance of $Cs_2O$ is thermal conductivity. It consists of lattice thermal conductivity (κL) and electrical thermal conductivity (κe) attributed to electrons and phonons, respectively. However, the BoltzTrap code cannot calculate lattice thermal conductivity (κL), so only electronic thermal conductivity is shown. Fig. 7(c) illustrates electronic thermal conductivity variation over the constant relaxation time (κe/τ) within the range of 100–1200 K for $Cs_2O$. As depicted in Fig. 7(c) the κe/τ increases with increase in temperature and reaches upto a maximum value of 0.92 x $10^{15}$W/mKs.

**Power factor and figure of merit**

To understand $Cs_2O$ thermoelectric properties, we calculated the power factor over the constant relaxation time (S²σ/τ) and the electronic figure of merit (ZeT = S²σT/κe) for $Cs_2O$. Fig. 7(d) shows the variation of S²σ/τ with temperature in the range of 200–1500 K. As depicted in the plot, the power factor increases to a maximum value of 7x$10^{10}$ W/Km$^{-1}$s$^{-1}$ at 800 K and then gradually decreases at higher temperatures.

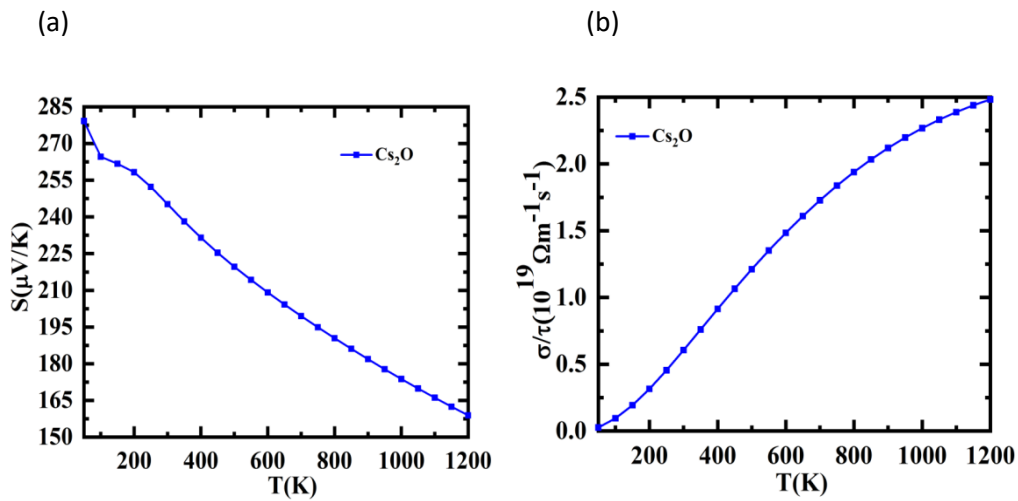

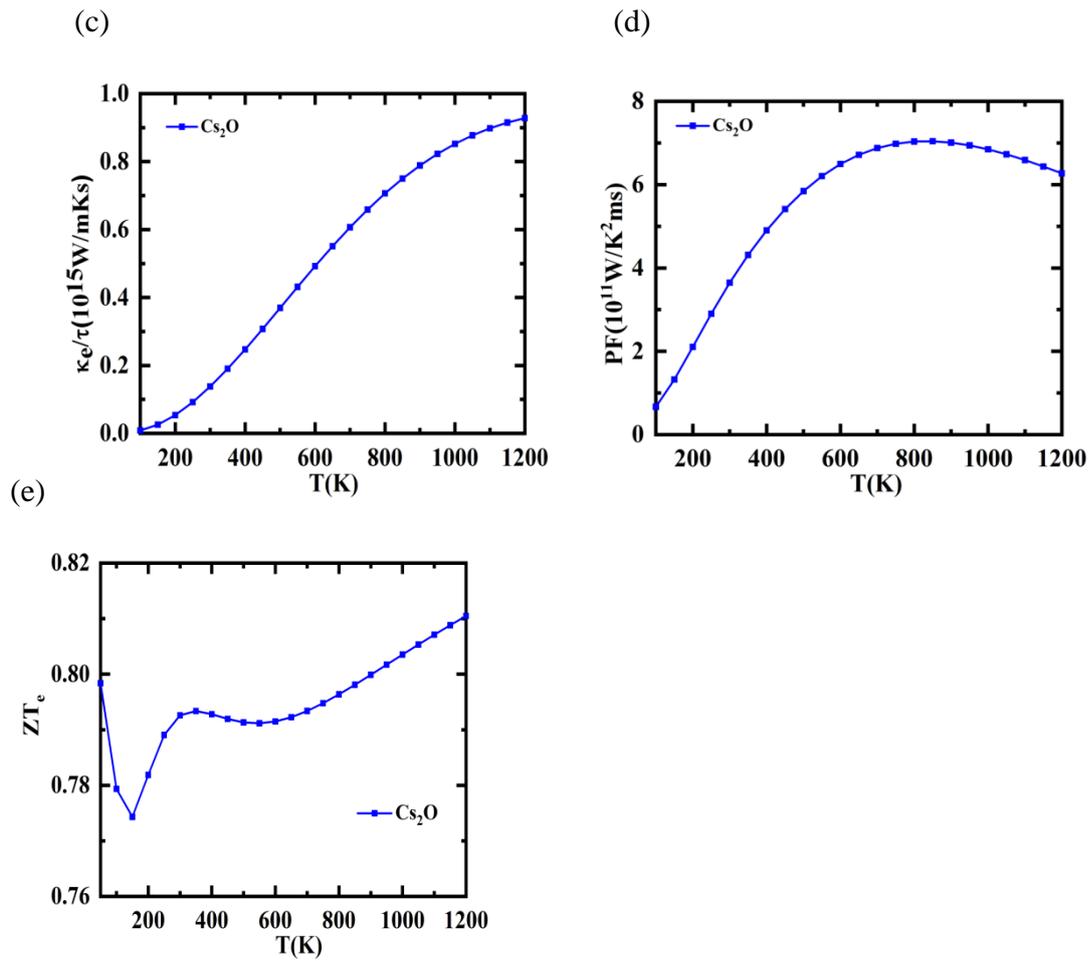

Fig. 8. Transport properties of $Cs_2O$ compound (a) Seebeck coefficient (b) Power factor (c) figure of merit (d) Electrical conductivity ($\sigma$) (e) Thermal conductivity (k).

Figure 8 depicts the temperature-dependent dimensionless figure of merit. With the increase in temperature the figure of merit increases. $Cs_2O$ has a value of 0.795 at room temperature, and this value increases further and reaches a maximum value of 0.810 at approximately 1200 K.

**Table 1.** Calculated lattice parameter a (in ˚A), bulk modulus $B0$ (in GPa) and its pressure derivatives $B$ for $Cs_2O$ compound

|  | Present work | |
|---|---|---|
| $Cs_2O$ | LDA | PBE-GGA |
| a(Å) | 6.92 | 7.31 |
| V(a.u)$^3$ | 560.34 | 659.10 |
| B(GPa) | 33.90 | 17.78 |
| B'' | 5.49 | 5.62 |
| $E_0$(Ry) | -31286.45 | -31311.69 |
| $E_{coh}$ | 3.38 | 3.45 |
| $E_{for}$ | -1.010 | -1.019 |

**Table 2.** Calculated optical properties of $Cs_2O$

|  | Present work | | |
|---|---|---|---|
| $Cs_2O$ | LDA | PBE-GGA | mBJ-GGA |
| Γ-Γ | 2.02 | 2.27 | 4.07 |
| X-Γ | 1.27 | 1.53 | 3.79 |
| X-X | 1.69 | 1.85 | 3.23 |
| ε(0) | 2.56 |  |  |
| n(0) | 1.59 |  |  |
| R(0) | 0.05 |  |  |

## 3. CONCLUSION

We conducted a thorough study of the electronic, structural, thermoelectric and optical characteristics of $Cs_2O$ using first-principle FP-APW+lo method within LDA, PBE-GGA, and mBJ-GGA approximations. Notably, the calculated band gaps within the mBJ-GGA approximation exhibit a significant enhancement compared to values obtained from LDA and PBE-GGA methods. Our analysis of the band structure confirms Cs$_2$O as an indirect band gap material, consistent with previous studies. Additionally, we investigated the critical point structure of the frequency-dependent complex dielectric function to identify optical transitions. To our knowledge, previous studies have not delved into the electronic structure and imaginary part of the dielectric constant for this compound. Thus, we believe our calculations can fill this gap in data and provide valuable insights into Cs$_2$O's properties.